\documentclass[%
 reprint,
 amsmath,amssymb,
 aps
]{revtex4-1}

\usepackage{graphicx}
\usepackage{dcolumn}
\usepackage{bm}
\usepackage{ifthen}
\usepackage{amssymb} 

\usepackage{float}

\newcommand{\Xc}{\Xi_{cc}^+}
\newcommand{\Lc}{\Lambda_c^+}

\begin{document}


\title{On the production properties of the Doubly-Charmed Baryons}

\author{Sergey Koshkarev}

\date{\today}

\begin{abstract}
This paper focuses on disagreement between theoretical predictions and experimental results of the production properties of Doubly Charmed Baryons. The kinematic dependencies are used to clarify the discrepancy between the SELEX data and the theory.  The production ratio of the $\Xc$ baryon in the SELEX kinematic region is presented. The recent experimental results are reviewed.\begin{description}
\item[PACS numbers]
14.65.Dw, 14.20.Lq, 13.60.Rj
\end{description}
\end{abstract}

\maketitle

\section{Introduction}

In 2002 the SELEX collaboration published the first observation of $\Xc$ baryon in the charged decay mode $\Xc \to \Lc K^- \pi^+$ from $\Lc \to p K^- \pi^+$ (1630 events) sample~\cite{SELEX2002}. In 2005 the SELEX collaboration reported an observation of $\Xc \to p D^+ K^-$ decay mode from 1450 $D^+ \to K^- \pi^+ \pi^+$ decays to complement the previously reported decay~\cite{SELEX2005}. The mass and lifetime also have been measured by SELEX (see Table~\ref{tab:table1}). 
\begin{table}[H]
 \caption{\label{tab:table1}
 	This table summarizes the SELEX results on measurements of the mass and lifetime of $\Xc$ baryon.
	}
	\begin{ruledtabular}
\begin{tabular}{ccc}
Source&$\Xc$ mass (MeV/$\text{c}^2$)&$\Xc$ lifetime (fs) \\
\hline
SELEX~\cite{SELEX2002}&$3519 \pm 1$&$<33$ at 90\% C.L. \\
SELEX~\cite{SELEX2005}&$3518 \pm 3$& not reported \\
\end{tabular}
\end{ruledtabular}
\end{table}
The production properties of $\Xc$ baryon can be obtained from the measurements and compared to that of $\Lc$ baryon and $D^+$ meson:
\[
R_{\Lc} = \frac{\sigma(\Xc) \cdot Br(\Xc \to \Lc K^- \pi^+)}{\sigma(\Lc)} \approx 0.015
\]
and
\[
R_{D^+} = \frac{\sigma(\Xc) \cdot Br(\Xc \to p D^+ K^-)}{\sigma(D^+)} \approx 0.004
\]
in kinematic region $x_F  > 0.3$~\cite{Mattson}. Using known fragmentation ratio $f(c \to \Lc) = 0.071 \pm 0.003~\text{(exp.)}\pm 0.018~\text{(br.)}$~\cite{BaBarLambda}  and assuming $Br(\Xc \to \Lc K^- \pi^+) \approx Br(\Lc \to p K^- \pi^+) =  (5.0 \pm 1.3)~\%$~\cite{PDG}, one can obtain the ratio of the production cross-section: \\
\begin{eqnarray*}
\frac{\sigma(\Xc)}{\sigma(c\bar{c})} = R_{\Lc} \cdot \frac{f(c \to \Lc)}{Br(\Xc \to \Lc K^- \pi^+)} \\ \simeq 2.1 \cdot 10^{-2}
\end{eqnarray*}
Comparing this result with theoretically predicted $\sigma(\Xc) / \sigma(c\bar{c}) \sim 10^{-6} - 10^{-5}$~\cite{KiselevUFN, Berezhnoy97}  production ratio for fixed-target experiments, we see that measured $\Xc$ production cross-section is at least $10^3$ times larger than theoretical prediction. This is a significant discrepancy between theory and experiment.
Recent searches in different production environments at BaBar~\cite{BaBar}, Belle~\cite{Belle} and LHCb~\cite{LHCb} experiments also have not shown evidence (see Table~\ref{tab:table2}) for the production properties reported by SELEX.
\begin{table*}
 \caption{\label{tab:table2}
 	This table shows the production properties on $\sigma(\Xc) \cdot Br(\Xc \to \Lc K^- \pi^+)/\sigma(\Lc)$ in different production environment. The LHCb results based on the assumption that $\Xc$ lifetime is 100 fs and 400 fs respectively.
	}
\begin{ruledtabular}
\begin{tabular}{lcccc}
Source&Upper Limit&C.L.&Production Environment&Integrated Luminosity\\
\hline
BaBar&$<2.7 \times 10^{-4}$&95\%&$e^-e^+$ collisions at $\sqrt{s}=10.58$ GeV&232 $\text{fb}^{-1}$\\
Belle&$<82-500$ fb&95\%&$e^-e^+$ collisions at $\sqrt{s}=10.58$ GeV&980 $\text{fb}^{-1}$\\
LHCb&$<1.5 \times 10^{-2}$&95\%&$pp$ collisions at  $\sqrt{s}=7$ TeV&0.65 $\text{fb}^{-1}$\\
LHCb&$<3.9 \times 10^{-4}$&95\%&$pp$ collisions at  $\sqrt{s}=7$ TeV&0.65 $\text{fb}^{-1}$
\end{tabular}
\end{ruledtabular}
\end{table*}
To clarify this issue we research kinematic dependencies of hadronic production properties of the doubly-charmed baryon.

\section{Production ratio of $\Xc$ baryon}
\subsection{SELEX}

The SELEX experiment is a fixed-target experiment used the Fermilab charged hyperon beam at 600 GeV/c to produce charm particles in a set of thin foil of Cu or in a diamond and operated in the $x_F > 0.3$ kinematic region. The negative beam composition was about 50\% $\Sigma^-$, 50\% $\pi^-$. The positive beam was 90\% protons. 

The Born approximation of the hadronic production of $\Xc$ baryon and the limitation of the model have been discussed in Ref.~\cite{Berezhnoy97}. Below we will use the results without any additional discussions. Following~\cite{Berezhnoy97} the parton level production cross-sections of $\Xc$ baryons:
\begin{eqnarray}
\label{eq:aaa}
\hat{\sigma}_{gg} = 213 \cdot \left( 1 - \frac{4 \cdot m_c}{\sqrt{\hat{s}}} \right)^{1.9}  \left( \frac{4 \cdot m_c}{\sqrt{\hat{s}}} \right)^{1.35}~~\text{pb},\\
\label{eq:bbb}
\hat{\sigma}_{q\bar{q}} = 206 \cdot \left( 1 - \frac{4 \cdot m_c}{\sqrt{\hat{s}}} \right)^{1.8}  \left( \frac{4 \cdot m_c}{\sqrt{\hat{s}}} \right)^{2.9}~~\text{pb}. 
\end{eqnarray}
One has to mention that numerical coefficients depend on the model parameters, so that $\hat{\sigma} \sim \alpha_s |R(0)|^2 / m_c^2$, where  $\alpha_s = 0.2$, $R(0) = 0.601 \text{GeV}^{3/2}$ and $m_c = 1.7$ GeV. Let us remind the reader that the hadron level cross-section can be presented as
\begin{equation}
\label{eq:ccc}
\sigma = \sum \int dx_1^{(k)} dx_2^{(m)} f_k (x_1, \mu) f_m (x_2, \mu) \hat{\sigma}(x_1, x_2),
\end{equation}
where $f_i (x, \mu)$ is a parton distribution function, $x$ is the ratio of the parton momentum to the momentum of the hadron and $\mu$ is the energy scale of the interaction.
Combining Eqs. \ref{eq:aaa}, \ref{eq:bbb}, \ref{eq:ccc} and using CTEQ6L~\cite{CTEQ6L} parametrization for parton distribution functions, we may expect $\Xc$ production cross-section in the kinematic region $x_F >0.4$~\cite{Mattson} to be
\[
\sigma(\Xc) \gtrsim 25~\text{pb}.
\]
Upon contracting this result with predictions of the Born approximation of two-quark hadronic production in the same kinematic region, $\Xc$ production ratio at the SELEX is
\[
\frac{\sigma(\Xc)}{\sigma(c\bar{c})} \sim (10^{-3} - 10^{-2})
\]

This result can be easily  interpreted taking into account to the four-quark production properties.
Assuming that $\hat{\sigma}(\Xc)$ is proportional to the parton level of the four-quark production cross-section $\hat{\sigma}(4Q)$ and following~\cite{Maltoni} we can see that $\hat{\sigma}(4Q)$ is growing up with the energy and turns out to be a leading process instead of two-quark production cross-section which dominates at the small energies. The uncertainty of the result came from the two-quark production cross-section. The SELEX analysis strategy~\cite{Mattson} requires $x_F > 0.4$ for a final state ($\Lc$ plus the positively charged track and the negatively charged track) which leads to uncertainty of the final charm $x_F$ distribution and cannot be considered without full modeling of the production environment.
\subsection{Belle and LHCb}
The Belle experiment  presented the upper limit on the $\sigma(e^+e^- \to \Xc X)$ is 82-500 fb for the decay mode with the $\Lc$ at $\sqrt{s} = 10.58$ GeV using 980 $\text{fb}^{-1}$. The most realistic calculations~\cite{KiselevUFN, Kiselev94} predict $\sigma(\Xc) \simeq (35 \pm 10)  \times 10^{-3}$ pb what turns out to be at least twice as less as the given limit.

Another recent result from the LHCb experiment provides the upper limits at 95\% C.L. on the ratio $\sigma(\Xc) \cdot Br(\Xc \to \Lc K^- \pi^+)/\sigma(\Lc)$ to be $1.5 \times 10^{-2}$ and $3.9 \times 10^{-4}$ for lifetimes 100 fs and 400 fs respectively, for an integrated luminosity of 0.65 $\text{fb}^{-1}$. It is compared with result from Ref.~\cite{KiselevUFN} $\sim 10^{-4} - 10^{-3}$. However, the LHCb did not reach the lifetime measured by the SELEX experiment yet.

\section{Conclusion}

In our paper we recalculated the hadronic production cross-section of the $\Xc$ baryon in the SELEX kinematic region and made the comparative analysis of experimental data and theoretical predictions. We found no significant discrepancy between the theory and the experimental results of the production properties of the doubly charmed baryons. 
\\
\acknowledgements{The authors would like to thank Dr. Alexander Rakitin for his friendly support and proofreading the manuscript.}


\begin{thebibliography}{99}

\bibitem{SELEX2002}
M.~Mattson {\sl et al.} (SELEX~Collaboration), Phys. Rev. Lett. 89,  112001 (2002).

\bibitem{SELEX2005}
A.~Ocherashvili {\sl et al.} (SELEX~Collaboration), Phys. Lett. B{\bf628}, 12-24 (2005).

\bibitem{Mattson}
M.~Mattson, Ph.D. thesis, Carnegie Mellon University, 2002.

\bibitem{BaBarLambda}
B.~Aubert {\sl et al.} (BaBar~Collaboration), Phys. Rev. D{\bf75}, 012003 (2007).

\bibitem{PDG}
J.~Beringer {\sl et al.}, Phys. Rev. D{\bf86}, 010001 (2012).

\bibitem{KiselevUFN}
V.~Kiselev and A.~Likhoded, Phys. Ups. {\bf45}, 455 (2002).

\bibitem{Berezhnoy97}
A.~Berezhnoy {\sl et al.}, Phys. Rev. D{\bf57}, 4385 (1998).

\bibitem{BaBar}
B.~Aubert {\sl et al.} (BaBar~Collaboration), Phys. Rev. D{\bf74}, 011103(R) 2006.

\bibitem{Belle}
Y.~Kato {\sl et al.} (Belle Collaboration),~Arxiv:1312.1026.

\bibitem{LHCb}
R.~Aaij {\sl et al.} (LHCb Collaboration), JHEP {\bf12}, 090 (2013).

\bibitem{CTEQ6L}
J.~Pumplin {\sl et al.}, JHEP {\bf07}, 012 (2002).

\bibitem{Maltoni}
J.~Andersen {\sl et al.}, JHEP {\bf11}, 061 (2004).

\bibitem{Kiselev94}
V.~Kiselev, A.~Likhoded, M.~Shevlyagin, Phys. Lett. B{\bf332}, 411-414 (1994).

\end{thebibliography}
\end{document}